\documentclass[prl,
twocolumn,
showpacs,aps]{revtex4}
\usepackage{times}
\usepackage{graphicx}
\usepackage{natbib}
\begin{document}
\title{Quantum Beat of Two Single Photons}
\author{Thomas Legero}
\author{Tatjana Wilk}
\author{Markus Hennrich}
\author{Gerhard Rempe}
\author{Axel Kuhn}
\affiliation {Max-Planck-Institut f\"ur Quantenoptik, Hans-Kopfermann-Str. 1, D-85748 Garching, Germany}
\date{\today}
\begin{abstract}
The interference of two single photons impinging on a beam splitter is measured in a time-resolved manner. Using long photons of different frequencies emitted from an atom-cavity system, a quantum beat with a visibility close to 100\% is observed in the correlation between the photodetections  at the output ports of the beam splitter. 
The time dependence of the beat amplitude reflects the coherence properties of the photons. Most remarkably, simultaneous photodetections  are never observed, so that  a temporal filter allows one  to obtain perfect two-photon coalescence even for non-perfect photons. 
\end{abstract}
\maketitle

The quantum nature of light impressively manifests itself in the fourth-order interference of two identical and mutually coherent single photons that impinge simultaneously on a beam splitter (BS). The photons coalesce and both leave the beam splitter in the same direction. Hong et al. first demonstrated this phenomenon with photon pairs from parametric down conversion \cite{Hong87} and Santori et al.  used the same effect to show the indistinguishability of independently generated photons that are successively emitted from a quantum dot embedded in a micro cavity \cite{Santori02}. In all experiments performed so far, the photons were short compared to the time resolution of the employed detectors, so that interference phenomena were only observed as a function of the spatial delay between the interfering photons \cite{Ou88}. 

To investigate the temporal dynamics behind this interference phenomenon, we now use an adiabatically driven strongly coupled atom-cavity system as single-photon emitter \cite{Hennrich00,Kuhn02:2,Kuhn02,Kuhn03}. Photons are generated by a unitary process, so that their temporal and spectral properties can be arbitrarily adjusted. 
In fact, the duration of the photons used in our experiment
 exceeds the time resolution of the employed single-photon counters by three orders of magnitude. This allows for  the first time an experimental investigation of fourth-order interference phenomena in a time-resolved manner  with photons  arriving simultaneously at the beam splitter \cite{Legero03}. 
We find perfect interference even if the frequency difference between the two photons exceeds their bandwidths. This surprising result is very robust against all kinds of fluctuations and opens up new possibilities in all-optical quantum information processing \cite{Knill01}.

\begin{figure}
\centering\includegraphics[width=7cm]{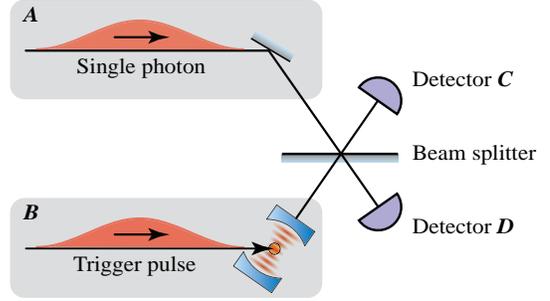}
\caption{\textbf{Fourth-order interference.} Single photons emerge from $A$ and $B$ and impinge simultaneously on a beam splitter. The photons  are so long that they give rise to distinct photodetections. The first detection projects the system into a superposition, which determines the probability of detecting the second photon with either one or the other detector. \label{Prinzip}}
\end{figure}

The principal scheme of the experiment  is sketched  in Fig.\,\ref{Prinzip}. We consider an initial situation where two single photons in modes $A$ and $B$ impinge simultaneously on a BS. In front of the BS, we distinguish states  $|1_{A,B}\rangle$ and $|0_{A,B}\rangle$, where either a photon is present or where it has been annihilated by transmission through the BS and subsequent detection by detector $C$ or $D$. Mode $A$ is an extended spatiotemporal photonic field mode, traveling  along an optical fiber, which initially carries a photon. The photon in mode $B$ emerges from a strongly coupled atom-cavity system, which is driven in a way that the photon is deterministically generated  by a vacuum-stimulated Raman transition between two long-lived atomic states  \cite{Hennrich00,Kuhn02}. In particular, the photon emitted from $B$ matches the photon from $A$. The initial  state of the total system, $A$ and $B$, is given by the  product state $|\Psi_{i}\rangle=|1_{A}1_B\rangle$. The effect of a first photodetection in the output mode $C$ or $D$ at time $t_{0}$ is evaluated by applying the respective photon annihilation operator, $\hat a_{C}$ or $\hat a_{D}$,  to $|\Psi_{i}\rangle$. The two operators behind the BS are linked to the two operators before the BS  by the unitary relation 
\begin{equation}
\hat a_{C,D}=(\hat a_{B}\pm \hat a_{A})/\sqrt{2},
\label{ulink}
\end{equation}
where $\hat a_{A}$ and $\hat a_{B}$ are operators that remove one  photon from $A$ and $B$, respectively. As the detection reveals no which-way information, the system is projected into one of the two superposition states,
\begin{equation}
|\Psi_{\pm}(t_{0})\rangle=\hat a_{C,D}|1_{A}1_{B}\rangle =  (|1_{A}, 0_{B}\rangle\pm|0_{A}, 1_{B}\rangle)/\sqrt{2},
\end{equation}
depending on which detector clicks. 

The  initial purity of the superposition, i.e. the balance between its two parts and their phase coherence, and, hence,  also the mutual coherence time of the interfering photons,  can now be measured by monitoring the time evolution of $|\Psi_{\pm}\rangle$. Moreover, the superposition can be systematically varied by controlling the relative phase between its two parts. The latter is achieved by introducing a small frequency difference between the photons from $A$ and $B$. Assume that the photon in the fiber has a frequency difference $\Delta$ with respect to the photon from the cavity. In this case, the two components of $|\Psi_{\pm}\rangle$ evolve with different frequencies, so that after a time  $\tau$, the two states have acquired a phase difference $\Delta \tau$. The new state then reads
\begin{equation}
|\Psi_{\pm}(t_0+\tau)\rangle = (|1_{A},0_B\rangle\pm e^{i\Delta \tau}|0_{A},1_B\rangle)/\sqrt{2}.
\label{interference}
\end{equation}
This state can be monitored by  photodetections. The probability to count a second photon with either detector $C$ or $D$, as a function of the relative phase $\Delta \tau$, reads
\begin{eqnarray}
\langle\Psi_{\pm}|\hat a^\dag_{C}\hat a^{}_{C}|\Psi_{\pm}\rangle &=&
\frac{1}{2}(1\pm\cos \Delta\tau) \quad\mbox{and}\nonumber\\
\langle\Psi_{\pm}|\hat a^\dag_{D}\hat a^{}_{D}|\Psi_{\pm}\rangle &=&
\frac{1}{2}(1\mp\cos \Delta\tau).\label{beat}
\end{eqnarray}
In a photon correlation experiment, a frequency difference between the interfering photons therefore  results in  a quantum beat signal in the correlation function that oscillates with frequency $\Delta$.  Moreover, Eq.\,(\ref{beat}) implies that the first and the second photon hit the same detector for $\Delta=0$. This corresponds to the well-known behavior of two indistinguishable photons that impinge simultaneously on a BS \cite{Hong87,Santori02}. However, in the present case, the two photodetections can have a time delay that can be as large as the  duration of the interfering photons.  Nevertheless, perfect two-photon coalescence is expected. Another remarkable consequence from Eq.\,(\ref{interference})  is that the initial phase difference between its two parts, induced by the first photodetection, is either $0$ or $\pi$.  Therefore the beat starts to oscillate at zero with the detection of a first photon, so that the cross-correlation function between the two BS output ports shows fringes with a visibility of 100\%. This distinguishes the present situation dramatically from the situation of two interfering coherent fields with frequency difference $\Delta$ that are superposed on a BS. In the latter case, the cross-correlation function would oscillate with a fringe visibility not exceeding 50\%, since photodetections do not influence the relative phase of the coherent fields. The present scheme is also different from other experiments, where quantum state reduction has been observed in optical cavity QED   \cite{Rempe91,Foster00}. In these experiments, the detection of a photon changes the state of a single system, whereas in the present case, the relative phase of two distinct modes, $A$ and $B$, is determined. 
 
We emphasize that the above way of calculating joint photodetection probabilities is strongly simplified. 
A more detailed analysis, that comes to the same conclusions, can be found in Ref. \cite{Legero03}.

\begin{figure}
\centering\includegraphics[width=6.9cm]{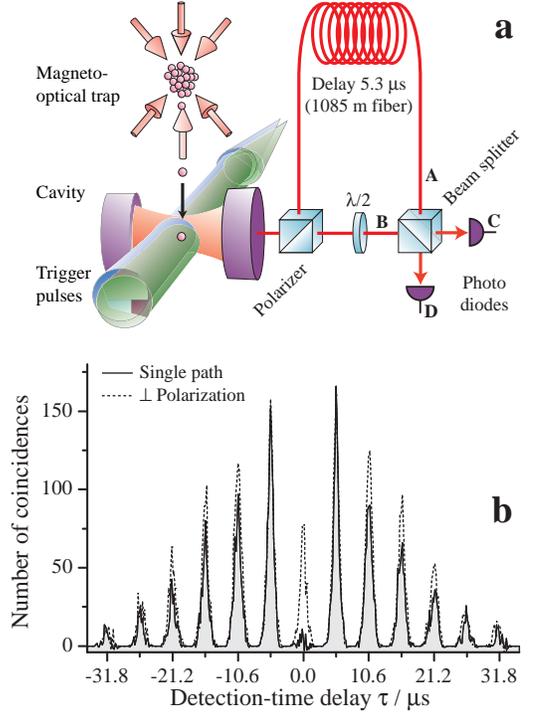}\caption{\textbf{Atoms and Photons. (a):} Triggered by laser pulses, an atom-cavity system emits unpolarized single photons. They are randomly directed by a polarizing beam splitter along two paths towards a non-polarizing beam splitter (BS). A photon traveling along path $A$ gets delayed  so that it impinges on the BS simultaneously with a subsequent photon that travels along path $B$. \textbf{(b):} Number of coinciding  photodetections in the two output ports  as a function of the time difference between the detections: If only a single path is open, a  Hanbury-Brown \& Twiss measurement of the  intensity correlation is performed, showing antibunching (solid line). If both  paths are open but have perpendicular polarization, no interference takes place and the BS randomly directs the photons to $C$ and $D$. This leads to coincidences at $\tau\approx 0$ (dashed line). All traces result from a convolution with a 48\,ns wide square time-bin function. 
\label{HOMandDELAY}}
\end{figure}

The experimental  setup is  sketched in Fig.\,\ref{HOMandDELAY}a. 
$^{85}$Rb atoms released from a magneto-optical trap fall with $2\,$m/s through a cavity of Finesse $F=60000$ with $(g_{max},\kappa,\gamma)/2\pi =  (3.1, 1.25, 3.0)\,$MHz, where $g_{max}$ is the optimal atom-cavity coupling constant, and $\kappa$ and $\gamma$ are the field decay rates of cavity and atom. The atoms enter one-at-a-time with a probability that is 66 times higher than the probability of having more atoms. Each atom is prepared in $|e\rangle \equiv |5S_{1/2}, F=3\rangle$, while the cavity is resonant with the transition between $|g\rangle\equiv |5S_{1/2}, F=2\rangle$ and $|x\rangle\equiv |5P_{3/2}, F=3\rangle$. The atom experiences a sequence of laser pulses that alternate between triggering single-photon emissions and repumping the atom to state $|e\rangle$: The $2\,\mu$s long trigger pulses are resonant with the $|e\rangle\leftrightarrow |x\rangle$ transition and their Rabi frequency increases linearly to $\Omega_{max}/2\pi=17.8\,$MHz. In connection with the vacuum-field of the cavity stimulating the $|x\rangle\leftrightarrow |g\rangle$ transition, these pulses drive an adiabatic passage (STIRAP) to $|g\rangle$. This transition goes hand-in-hand with a photon emission. Between two emissions, another laser pumps the atom from $|g\rangle$ to $|x\rangle$, from where it decays back to $|e\rangle$. This is complemented by a $\pi$-polarized laser driving the transition $|5S_{1/2}, F=3\rangle \leftrightarrow |5P_{3/2}, F=2\rangle$ to produce  a high degree of spin-polarization in $m_{F}=\pm 3$, with a large coupling to the cavity. To discard the photons emerging during this process,  the detectors are electronically gated. This leads to a modulation of the dark-count rate and, hence, to a triangular modulation of the background contribution to all correlation functions measured with detectors $C$ and $D$, with maxima showing an average number of 3.2 correlations/48\,ns. All data shown here have been corrected for this periodic background. 

We now consider the case where the atom-cavity system emits two  photons, one-after-the-other, with a time separation of $5.3\,\mu$s, deliberately introduced by the periodicity of our trigger pulse sequence. We suppose that the first photon travels along an optical fiber (mode $A$) and hits a 50:50 BS at the fiber output at the same time as the second photon, provided the latter comes directly from the cavity (mode $B$). To characterize the system, we first close the fiber, so that photons impinge only in mode $B$. For this situation, Fig.\,\ref{HOMandDELAY}b shows the intensity correlation function, measured with  detectors $C$ and $D$,  as a function of the time difference, $\tau$, between photodetections as a solid line. The central peak is missing, i.e. the light shows strong antibunching and photons are emitted one-by-one \cite{Kuhn02}. Next, both paths to the BS are opened, so that photons can impinge  simultaneously on the BS. Interference is suppressed by adjusting the  $\lambda/2$ retardation plate at input port $B$ so that the two light fields are polarized perpendicular to each other (dashed line). In this case, each photon is randomly directed onto one of the detectors, $C$ or $D$. This leads to a non-vanishing correlation signal  at $\tau\approx 0$, which is a factor of two smaller than the neighboring peaks at $\tau=\pm 5.3\,\mu$s. The central peak has a duration of $640\,$ns (half width at $\frac{1}{e}$ maximum), which comes from the convolution of two 450\,ns long single photons (half width at $\frac{1}{e}$ maximum of the intensity). In the following, the  signal obtained for perpendicularly polarized photons  is used as a reference, since any interference leads to a significant deviation. Note that all the experimental traces presented here are not sensitive to photon losses, since only measured coincidences contribute. Moreover, they can be compared without normalization, since data was always recorded until 980 coincidences were obtained in the correlation peaks at $\tau=\pm 5.3\,\mu$s. This required to load and release atoms from the magneto-optical trap about $10^5$ times.

\begin{figure}
\centering\includegraphics[width=7.5cm]{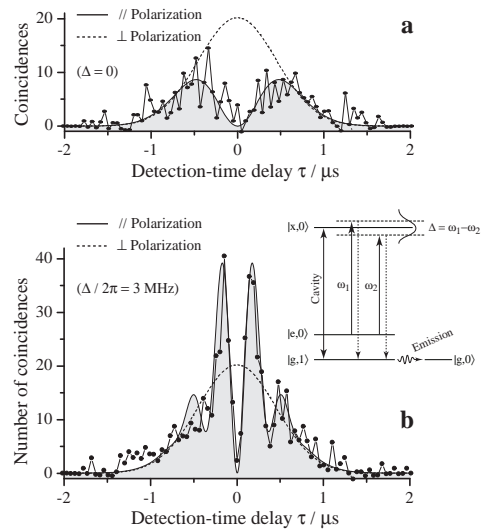}
\caption{\textbf{Quantum Beat.} Number of coinciding photodetections as a function of the time difference, $\tau$, between  the detections (only the central peak is shown, see Fig.\,\ref{HOMandDELAY}b). Both paths are open and have parallel polarization (circles). The solid lines represent a numerical fit  to the data \protect\cite{Legero03}.  A Gaussian fit to the reference signal (perpendicular polarization, dashed line) is also shown. 
\textbf{(a):} Photons of identical frequencies lead to a 460\,ns wide central minimum. This lack of coincidences is caused by coalescing photons that leave the BS through the same port. Depth and width of the minimum indicate the initial purity of the superposition and the mutual coherence time of the photons, respectively. \textbf{(b):} The atom-cavity system is driven by a sequence of laser pulses with a frequency difference $\Delta=|\omega_{1}-\omega_{2}|=2\pi\times 3\,$MHz between consecutive pulses (see level scheme). This gives rise to a frequency difference between consecutive photon emissions, which leads to a quantum beat in the correlation function starting at $\tau=0$. 
\label{QuantumBeat}}
\end{figure}

Experimental results obtained for parallel polarization are displayed in Fig.\,\ref{QuantumBeat}a. The photons interfere and the first photodetection reveals  no which-way information. Therefore the system is projected into the superposition state $|\Psi_{\pm}\rangle$. In contrast to the expectations from the above discussion, the correlation signal does not vanish completely, in particular for non-zero detection time delay. Instead, a pronounced  minimum  is observed around $\tau=0$. We interpret the depth of this minimum as a measure of the initial purity of the superposition state,  and we attribute its limited  width to the average mutual dephasing of the interfering photons (see below).  Moreover, as shown in Fig.\,\ref{QuantumBeat}b, we resolve a pronounced oscillation of the correlation function, starting with a minimum at $\tau=0$,  when a frequency difference of $\Delta/2\pi=3\,$MHz is introduced between the interfering photons. The first maxima of the oscillation are found at $|\tau|\approx\pi/\Delta$, where the two parts of $|\Psi_{\pm}\rangle$ have acquired a phase difference of  $\pm\pi$. If the photons are detected with this time difference,  they are registered by different detectors and give rise to a coincidence count. Therefore the number of coincidences in these maxima exceeds the reference level, measured with perpendicular polarization, by a factor of two. This underpins the phase coherence of the whole process and shows that it is possible to arbitrarily adjust the relative phase between the two parts of $|\Psi_{\pm}\rangle$. The initial purity of the superposition and the balance between its two parts is characterized by the visibility of the beat signal at $\tau=0$. This visibility exceeds 90\%, indicating that the superposition is nearly perfect. 

The mutual coherence  time of the interfering photons is obtained from the damping of the quantum beat or, alternatively, from the width of the  two-photon interference dip. In both cases, a  coherence time of 460\,ns (half width at $\frac{1}{e}$ dip-depth) is observed, which exceeds the 64\,ns decay time of the cavity, as well as the 27\,ns lifetime of the atom's excited state. Hence,   the intrinsic lifetimes do not limit the coherence. However, for perfectly transform limited photons, one would expect to see no decrease of the quantum-beat visibility for $\Delta\neq 0$, and no correlation at all for $\Delta=0$. This is obviously not the case --  a numerical fit to the measured data based on an  analytical model \cite{Legero03}  (solid lines in Fig.\,\ref{QuantumBeat}) shows that the observed coherence time can be explained by an inhomogeneous broadening of $\delta\omega/2\pi=690\,$kHz, which exceeds the 350\,kHz bandwidth of transform limited photons. No specific broadening mechanism could be identified, and therefore we attribute this to several  technical reasons: static and fluctuating magnetic fields affect the energies of the magnetic substates and spread the photon frequencies over a range of 160\,kHz, and the trigger laser has a linewidth of 50\,kHz which is mapped to the photons. Moreover, diabatically generated photons  lead to an additional broadening.

To summarize, we have observed the fourth-order interference of two individual photons impinging on a beam splitter in a time-resolved manner. With photons of different frequencies, a quantum beat is found in the correlation between the photodetections at the output ports of the beam splitter. This beat oscillates with the frequency difference of the interfering photons. The interference fringes are only visible for  photons that are detected within their mutual coherence time. Moreover, our measurements  reveal that identical photons coalesce, i.e. they leave the beam splitter as a pair, provided they do not dephase with respect to each other. Any deviation from perfect coalescence, observed for non-zero detection-time delay, can be attributed to  a random dephasing due to an inhomogeneous broadening of the photon spectrum. We therefore conclude that a temporal filter, which only accepts time intervals between photodetections shorter than the mutual coherence time, is a way to obtain nearly perfect two-photon interference, even if the coherence properties of the photons are not ideal. This makes linear optical quantum computing \cite{Knill01} much more feasible with todays technology. 

Moreover, we point out that the present experiment is formally equivalent to a setup composed of two independent atom-cavity systems, since the photon traveling along the optical fiber could as well be released directly from an independent (second) atom-cavity system. Provided the time the photons need to travel from the cavities to the detectors is much shorter than their mutual coherence time (as is in fact the case for mode $B$ in our experiment), the first photodetection would establish an entanglement between the distant atom-cavity systems \cite{Cabrillo99,Hong02,Feng03,Browne03}, since the states $|1_{A,B}\rangle$ and $|0_{A,B}\rangle$ refer in this case to these systems. This entanglement would live until it is destructively probed by a second photodetection. Our results therefore pave the way towards distributed quantum computing and teleportation of atomic quantum states  \cite{Bose99,Lloyd01}.  

\begin{acknowledgements}
This work was supported by the focused research program `Quantum Information Processing' and the SFB 631 of the Deutsche Forschungsgemeinschaft, and by the European Union through the IST (QUBITS, QGATES) and IHP (QUEST, CONQUEST) programs.
\end{acknowledgements}


\begin{thebibliography}{16}
\expandafter\ifx\csname natexlab\endcsname\relax\def\natexlab#1{#1}\fi
\expandafter\ifx\csname bibnamefont\endcsname\relax
  \def\bibnamefont#1{#1}\fi
\expandafter\ifx\csname bibfnamefont\endcsname\relax
  \def\bibfnamefont#1{#1}\fi
\expandafter\ifx\csname citenamefont\endcsname\relax
  \def\citenamefont#1{#1}\fi
\expandafter\ifx\csname url\endcsname\relax
  \def\url#1{\texttt{#1}}\fi
\expandafter\ifx\csname urlprefix\endcsname\relax\def\urlprefix{URL }\fi
\providecommand{\bibinfo}[2]{#2}
\providecommand{\eprint}[2][]{\url{#2}}

\bibitem[{\citenamefont{Hong et~al.}(1987)\citenamefont{Hong, Ou, and
  Mandel}}]{Hong87}
\bibinfo{author}{\bibfnamefont{C.~K.} \bibnamefont{Hong}},
  \bibinfo{author}{\bibfnamefont{Z.~Y.} \bibnamefont{Ou}}, \bibnamefont{and}
  \bibinfo{author}{\bibfnamefont{L.}~\bibnamefont{Mandel}},
  \bibinfo{journal}{Phys. Rev. Lett.} \textbf{\bibinfo{volume}{59}},
  \bibinfo{pages}{2044} (\bibinfo{year}{1987}).

\bibitem[{\citenamefont{Santori et~al.}(2002)\citenamefont{Santori, Fattal,
  Vu\v{c}kovi\'{c}, Solomon, and Yamamoto}}]{Santori02}
\bibinfo{author}{\bibfnamefont{C.}~\bibnamefont{Santori}},
  \bibinfo{author}{\bibfnamefont{D.}~\bibnamefont{Fattal}},
  \bibinfo{author}{\bibfnamefont{J.}~\bibnamefont{Vu\v{c}kovi\'{c}}},
  \bibinfo{author}{\bibfnamefont{G.~S.} \bibnamefont{Solomon}},
  \bibnamefont{and} \bibinfo{author}{\bibfnamefont{Y.}~\bibnamefont{Yamamoto}},
  \bibinfo{journal}{Nature} \textbf{\bibinfo{volume}{419}},
  \bibinfo{pages}{594} (\bibinfo{year}{2002}).

\bibitem[{\citenamefont{Ou et~al.}(1988)\citenamefont{Ou, and
  Mandel}}]{Ou88}
 \bibinfo{author}{\bibfnamefont{Z.~Y.} \bibnamefont{Ou}}, \bibnamefont{and}
  \bibinfo{author}{\bibfnamefont{L.}~\bibnamefont{Mandel}},
  \bibinfo{journal}{Phys. Rev. Lett.} \textbf{\bibinfo{volume}{61}},
  \bibinfo{pages}{54} (\bibinfo{year}{1988}).

\bibitem[{\citenamefont{Kuhn et~al.}(2002)\citenamefont{Kuhn, Hennrich, and
  Rempe}}]{Kuhn02}
\bibinfo{author}{\bibfnamefont{A.}~\bibnamefont{Kuhn}},
  \bibinfo{author}{\bibfnamefont{M.}~\bibnamefont{Hennrich}}, \bibnamefont{and}
  \bibinfo{author}{\bibfnamefont{G.}~\bibnamefont{Rempe}},
  \bibinfo{journal}{Phys. Rev. Lett.} \textbf{\bibinfo{volume}{89}},
  \bibinfo{pages}{067901} (\bibinfo{year}{2002}).

\bibitem[{\citenamefont{Hennrich et~al.}(2000)\citenamefont{Hennrich, Legero,
  Kuhn, and Rempe}}]{Hennrich00}
\bibinfo{author}{\bibfnamefont{M.}~\bibnamefont{Hennrich}},
  \bibinfo{author}{\bibfnamefont{T.}~\bibnamefont{Legero}},
  \bibinfo{author}{\bibfnamefont{A.}~\bibnamefont{Kuhn}}, \bibnamefont{and}
  \bibinfo{author}{\bibfnamefont{G.}~\bibnamefont{Rempe}},
  \bibinfo{journal}{Phys. Rev. Lett.} \textbf{\bibinfo{volume}{85}},
  \bibinfo{pages}{4872} (\bibinfo{year}{2000}).

\bibitem[{\citenamefont{Kuhn and Rempe}(2002)}]{Kuhn02:2}
\bibinfo{author}{\bibfnamefont{A.}~\bibnamefont{Kuhn}} \bibnamefont{and}
  \bibinfo{author}{\bibfnamefont{G.}~\bibnamefont{Rempe}}, in
  \emph{\bibinfo{booktitle}{Experimental Quantum Computation and Information}},
  edited by \bibinfo{editor}{\bibfnamefont{F.~D.} \bibnamefont{Martini}}
  \bibnamefont{and} \bibinfo{editor}{\bibfnamefont{C.}~\bibnamefont{Monroe}}
  (\bibinfo{publisher}{IOS-Press}, \bibinfo{year}{2002}), vol.
  \bibinfo{volume}{148}, pp. \bibinfo{pages}{37--66}.

\bibitem[{\citenamefont{Kuhn et~al.}(2003)\citenamefont{Kuhn, Hennrich, and
  Rempe}}]{Kuhn03}
\bibinfo{author}{\bibfnamefont{A.}~\bibnamefont{Kuhn}},
  \bibinfo{author}{\bibfnamefont{M.}~\bibnamefont{Hennrich}}, \bibnamefont{and}
  \bibinfo{author}{\bibfnamefont{G.}~\bibnamefont{Rempe}}, in
  \emph{\bibinfo{booktitle}{Quantum Information Processing}}, edited by
  \bibinfo{editor}{\bibfnamefont{T.}~\bibnamefont{Beth}} \bibnamefont{and}
  \bibinfo{editor}{\bibfnamefont{G.}~\bibnamefont{Leuchs}}
  (\bibinfo{publisher}{Wiley-VCH, Berlin}, \bibinfo{year}{2003}), pp.
  \bibinfo{pages}{182--195}.

\bibitem[{\citenamefont{Legero et~al.}(2003)\citenamefont{Legero, Wilk, Kuhn,
  and Rempe}}]{Legero03}
\bibinfo{author}{\bibfnamefont{T.}~\bibnamefont{Legero}},
  \bibinfo{author}{\bibfnamefont{T.}~\bibnamefont{Wilk}},
  \bibinfo{author}{\bibfnamefont{A.}~\bibnamefont{Kuhn}}, \bibnamefont{and}
  \bibinfo{author}{\bibfnamefont{G.}~\bibnamefont{Rempe}},
  \bibinfo{journal}{Appl. Phys. B} \textbf{\bibinfo{volume}{77}},
  \bibinfo{pages}{797} (\bibinfo{year}{2003}).

\bibitem[{\citenamefont{Knill et~al.}(2001)\citenamefont{Knill, Laflamme, and
  Milburn}}]{Knill01}
\bibinfo{author}{\bibfnamefont{E.}~\bibnamefont{Knill}},
  \bibinfo{author}{\bibfnamefont{R.}~\bibnamefont{Laflamme}}, \bibnamefont{and}
  \bibinfo{author}{\bibfnamefont{G.~J.} \bibnamefont{Milburn}},
  \bibinfo{journal}{Nature} \textbf{\bibinfo{volume}{409}}, \bibinfo{pages}{46}
  (\bibinfo{year}{2001}).

\bibitem[{\citenamefont{Rempe et~al.}(1991)\citenamefont{Rempe, Thompson,
  Brecha, Lee, and Kimble}}]{Rempe91}
\bibinfo{author}{\bibfnamefont{G.}~\bibnamefont{Rempe}},
  \bibinfo{author}{\bibfnamefont{R.~J.} \bibnamefont{Thompson}},
  \bibinfo{author}{\bibfnamefont{R.~J.} \bibnamefont{Brecha}},
  \bibinfo{author}{\bibfnamefont{W.~D.} \bibnamefont{Lee}}, \bibnamefont{and}
  \bibinfo{author}{\bibfnamefont{H.~J.} \bibnamefont{Kimble}},
  \bibinfo{journal}{Phys. Rev. Lett.} \textbf{\bibinfo{volume}{67}},
  \bibinfo{pages}{1727} (\bibinfo{year}{1991}).

\bibitem[{\citenamefont{Foster et~al.}(2000)\citenamefont{Foster, Orozco,
  Castro-Beltran, and Carmichael}}]{Foster00}
\bibinfo{author}{\bibfnamefont{G.~T.} \bibnamefont{Foster}},
  \bibinfo{author}{\bibfnamefont{L.~A.} \bibnamefont{Orozco}},
  \bibinfo{author}{\bibfnamefont{H.~M.} \bibnamefont{Castro-Beltran}},
  \bibnamefont{and} \bibinfo{author}{\bibfnamefont{H.~J.}
  \bibnamefont{Carmichael}}, \bibinfo{journal}{Phys. Rev. Lett.}
  \textbf{\bibinfo{volume}{85}}, \bibinfo{pages}{3149} (\bibinfo{year}{2000}).

\bibitem[{\citenamefont{Cabrillo et~al.}(1999)\citenamefont{Cabrillo, Cirac,
  Garc{\'i}a-Fern{\'a}ndez, and Zoller}}]{Cabrillo99}
\bibinfo{author}{\bibfnamefont{C.}~\bibnamefont{Cabrillo}},
  \bibinfo{author}{\bibfnamefont{J.~I.} \bibnamefont{Cirac}},
  \bibinfo{author}{\bibfnamefont{P.}~\bibnamefont{Garc{\'i}a-Fern{\'a}ndez}},
  \bibnamefont{and} \bibinfo{author}{\bibfnamefont{P.}~\bibnamefont{Zoller}},
  \bibinfo{journal}{Phys. Rev. A} \textbf{\bibinfo{volume}{59}},
  \bibinfo{pages}{1025} (\bibinfo{year}{1999}).

\bibitem[{\citenamefont{Hong and Lee}(2002)}]{Hong02}
\bibinfo{author}{\bibfnamefont{J.}~\bibnamefont{Hong}} \bibnamefont{and}
  \bibinfo{author}{\bibfnamefont{H.-W.} \bibnamefont{Lee}},
  \bibinfo{journal}{Phys. Rev. Lett.} \textbf{\bibinfo{volume}{89}},
  \bibinfo{pages}{237901} (\bibinfo{year}{2002}).

\bibitem[{\citenamefont{Feng et~al.}(2003)\citenamefont{Feng, Zhang, Li, Gong,
  and Xu}}]{Feng03}
\bibinfo{author}{\bibfnamefont{X.-L.} \bibnamefont{Feng}},
  \bibinfo{author}{\bibfnamefont{Z.-M.} \bibnamefont{Zhang}},
  \bibinfo{author}{\bibfnamefont{X.-D.} \bibnamefont{Li}},
  \bibinfo{author}{\bibfnamefont{S.-Q.} \bibnamefont{Gong}}, \bibnamefont{and}
  \bibinfo{author}{\bibfnamefont{Z.-Z.} \bibnamefont{Xu}},
  \bibinfo{journal}{Phys. Rev. Lett.} \textbf{\bibinfo{volume}{90}},
  \bibinfo{pages}{217902} (\bibinfo{year}{2003}).

\bibitem[{\citenamefont{Browne et~al.}(2003)\citenamefont{Browne, Plenio, and
  Huelga}}]{Browne03}
\bibinfo{author}{\bibfnamefont{D.~E.} \bibnamefont{Browne}},
  \bibinfo{author}{\bibfnamefont{M.~B.} \bibnamefont{Plenio}},
  \bibnamefont{and} \bibinfo{author}{\bibfnamefont{S.~F.}
  \bibnamefont{Huelga}}, \bibinfo{journal}{Phys. Rev. Lett.}
  \textbf{\bibinfo{volume}{91}}, \bibinfo{pages}{067901}
  (\bibinfo{year}{2003}).

\bibitem[{\citenamefont{Bose et~al.}(1999)\citenamefont{Bose, Knight, Plenio,
  and Vedral}}]{Bose99}
\bibinfo{author}{\bibfnamefont{S.}~\bibnamefont{Bose}},
  \bibinfo{author}{\bibfnamefont{P.~L.} \bibnamefont{Knight}},
  \bibinfo{author}{\bibfnamefont{M.~B.} \bibnamefont{Plenio}},
  \bibnamefont{and} \bibinfo{author}{\bibfnamefont{V.}~\bibnamefont{Vedral}},
  \bibinfo{journal}{Phys. Rev. Lett.} \textbf{\bibinfo{volume}{83}},
  \bibinfo{pages}{5158} (\bibinfo{year}{1999}).

\bibitem[{\citenamefont{Lloyd et~al.}(2001)\citenamefont{Lloyd, Shahriar,
  Shapiro, and Hemmer}}]{Lloyd01}
\bibinfo{author}{\bibfnamefont{S.}~\bibnamefont{Lloyd}},
  \bibinfo{author}{\bibfnamefont{M.~S.} \bibnamefont{Shahriar}},
  \bibinfo{author}{\bibfnamefont{J.~H.} \bibnamefont{Shapiro}},
  \bibnamefont{and} \bibinfo{author}{\bibfnamefont{P.~R.}~\bibnamefont{Hemmer}},
  \bibinfo{journal}{Phys. Rev. Lett.} \textbf{\bibinfo{volume}{87}},
  \bibinfo{pages}{167903} (\bibinfo{year}{2001}).
\end{thebibliography}
 \end{document}